\documentstyle[12pt,aps]{revtex}
\begin{document}
\title{
WHAT CHIRAL SYMMETRY CAN TELL US ABOUT HADRON CORRELATORS IN MATTER
}
\author{B.V.Krippa\footnote{Talk given on BARYONS 98, 
Bonn, Sept. 22-26,1998. \\
e-mail boris@al20.inr.troitsk.ru}}
\address{Institute for Nuclear Research of
 the Russian Academy of Sciences, Moscow Region 117312,
Russia}

\maketitle

\begin{abstract}
The constraints imposed by chiral symmetry on hadron correlation
functions in  nuclear medium are discussed.
It is shown that these constraints
imply some  structure of the in-medium hadron correlators,
 lead to the cancellation of the order $\rho m_\pi$  term
in the in-medium nucleon correlator and result in the effect
of mixing of the chiral partners correlators, 
 reflecting the phenomena of partial restoration of chiral symmetry.
The different scenarios of such restorations are briefly discussed.
\end{abstract}

\section{Introduction}

There are little doubts that Chiral Symmetry (CS) is one of the
most important principles
of low-energy hadron physics. \cite {Ad}. In the limit
 of massless quarks the QCD 
Lagrangian is symmetric with respect 
to the $SU(N) \times SU(N)$ chiral group. It is generally believed
that this symmetry is spontaneously broken.
 CS breaking manifests itself in the absence
of chiral multiplets of the particles with the same masses but
different parities, for example, $\rho-a_1$ or $\sigma-\pi$
mesons. In the language of the correlation functions the broken chiral 
symmetry means that the lowest pole positions of vector and axial 
vector correlators, describing the $\rho$ and $a_1$ mesons
, are different. The restoration of the symmetry in vacuum would
 result in the identity of the corresponding
correlators which in turn leads to the same masses of the chiral partners.
 The other manifestation of the
 spontaneously broken CS is the occurrence of the nonzero order parameters
like two
quark condensate ${\overline qq}$. In the case of 
of hadron interactions in vacuum  $q{\overline {q}}$=0 would imply that
hadron masses become much smaller compared to its observed values.
 The relationships
between condensates, hadron masses and corresponding correlators  
are significantly more complicated in the presence of medium.
For example,  the change of the nucleon mass in
 medium is not completely determined by the corresponding change of 
the quark condensate.
The other example is the 
CS restoration for the correlators of the chiral partners. In matter
 the identity of
the corresponding in-medium correlators does not nesseseraly means the
degeneracy of the effective masses of chiral partners.
The identity of the masses of chiral partners at
 the point of restoration
is only one of the posibilities.
 The dynamics of hadrons in nuclear medium is
 described by the corresponding
correlators. Let's consider the case of the two-point correlators. It can 
be written in the form    
\begin{equation}
\Pi(p)=
i\!\int d^4x e^{ip\cdot x}\langle \Psi|T \{J(x)J(0)\}
|\Psi\rangle.
\end{equation}
Where $J$ is the interpolating current in the corresponding hadron channel
and the matrix element is taken over the ground states of the system with
finite density $\rho$. The position of the lowest pole
 as the function of density determines the in-medium mass 
of hadron.  If  wave function $\Psi$ describes the infinite system of non 
interacting nucleons than the above correlator reflects the dynamics
of the probe hadron moving in some mean field formed by nuclear matter.
To calculate the corrections to this  picture one needs to 
take into account the nuclear pions. Then the  correlator
 can represented  by the sum of two terms describing the contributions from the system of
noninteracting nucleons and pionic corrections.
 Let's consider the specific part of this corrections  where pion is 
emitted and then absorbed by nuclear matter. The corresponding piece of
the correlators can be represented in the form
\begin{equation}
\int {d\hbox{\bf k}\over 4\omega_k}
{d\hbox{\bf k}'\over \omega_k'}\langle\Psi|a^{a\dagger}_{\bf k}
a^b_{\bf k'}|\Psi\rangle
\,i\!\int d^4x e^{ip\cdot x}\langle \Psi\,\pi^a(\hbox{\bf k})|T\{J(x)
J(0)\}|\Psi\,\pi^b(\hbox{\bf k}')\rangle
\label{piinout}
\end{equation}
The sum over the isospin indices is assumed.
The terms with the different time orderings can be accounted for in a 
similar manner.
We consider the nuclear pions in the chiral limit. Since we are interested
in the properties of hadron correlators which are related to chiral symmetry
 treating nuclear pions in the chiral limit seems to be a reasonable 
assumption in our case. By using the  soft-pion
theorem 
 the part of the correlator describing
the pionic corrections can be written
as follows
\begin{equation}
\Pi^{\pi}={-i\over 2}\xi\int d^4x e^{ip\cdot x}
\langle \Psi|[Q^{a}_{5},[Q^{a}_{5},T\{J(x),J(0)\}]]
|\Psi\rangle,
\end{equation}
where we denoted 
 $\xi={\rho\overline{\sigma}_{\pi{\scriptscriptstyle N}}\over f_\pi^2 m_\pi^2}$
and 
 $\overline{\sigma}_{\pi{\scriptscriptstyle N}}$
is the leading nonanalytic  part of the pion-nucleon sigma term
$\sigma_{\pi{\scriptscriptstyle N}}$
The chiral expansion of $\sigma_{\pi{\scriptscriptstyle N}}$ reads as
\begin{equation}
{\sigma}_{\pi{\scriptscriptstyle N}}=A m_\pi^2 -
 {9\over 16}({g_{\pi N}\over 2M_N})^2m_\pi^3 + ....... 
\end{equation}
First, analytic term describes the short range contributions.
In contrast, the nonanalytic term is due to long distance contribution
of the pion cloud.
Let's consider the correlator of two nucleon interpolating current.
 Matter can influence the properties of 
the QCD vacuum and thus change the two quark condensate. It is usually
believed that the reduction of  $<\overline{q}q>$ is related
 to the effect of partial restoration of CS. 
To the first order in the density the evolution of the two-quark condensate
is given by \cite{Dr,Co}
\begin{equation}
\langle\Psi|\overline {q}q|\Psi\rangle=\langle 0|\overline {q}q|0\rangle
(1-{\sigma_{\pi{\scriptscriptstyle N}}\over f_\pi^2 m_\pi^2}\rho),
\label{ratio}
\end{equation}
It turned out that not any change of  
$\overline qq$ is in the one-to-one correspondence with the  
CS restoration phenomena \cite{Birse} since the terms  of order $ m_\pi$
presenting in the condensate are not allowed in in-medium nucleon mass.
  Let's
show how to cancel the terms not allowed by CS in QCD sum rules. In QCD
sum rules one relates the characteristics of  QCD vacuum and the 
phenomenological in-medium nucleon spectral density.
The effects of the low-momentum pions are long-ranged and
stem from the phenomenological
 representation of the
 in-medium nucleon correlator.
Assuming the Ioffe choice \cite{Iof} of the nucleon
interpolating current, making use
 of the transformation property of this current 
one can get the following chiral expansion of the in-medium nucleon 
correlator
\begin{equation}
\Pi(p)\simeq\Pi^{0}(p)-{\xi\over 2}\Bigl(\Pi^{0}(p)+\gamma_5 
\Pi^{0}(p)\gamma_5\Bigr),
\label{correxp}
\end{equation}
Where $\Pi^{0}(p)$ is the nucleon correlator in chiral limit.
It is useful to  decompose the correlator into three terms with the different
Dirac structures \cite{Co}
\begin{equation}
\Pi(p)=\Pi^{(s)}(p)+\Pi^{(p)}(p)p\llap/+\Pi^{(u)}(p)u\llap/,
\end{equation}
where $u^\mu$ is a unit four-vector  defining the rest-frame of nuclear
system.
 Only the piece $\Pi^{(s)}(p)$ gets affected by the chiral
corrections of order $\rho m_\pi$.
Splitting
the phenomenological expression of the nucleon correlator into pole and
 continuum parts one can get
\begin{equation}
\Pi(p)\simeq\Pi_{pole}(p)-{\xi\over 2}\gamma_5\Pi_{pole}(p)\gamma_5
+\left(1-{\xi\over 2}\right)\Pi^{0}_{cont}(p)
-{\xi\over 2}\gamma_5 \Pi^{0}_{cont}(p)\gamma_5,
\label{corrphen}
\end{equation}
Where we denoted $\Pi_{pole}(p)\simeq (1-{\xi\over 2})\Pi^{0}_{pole}(p)$
The explicit expression of the pole term has the following form
\cite{Co}
\begin{equation}
\Pi_{pole}(p)=-\lambda^{*2}{p\llap/+M^*+V\gamma_0\over 2E(\hbox{\bf p})
[p^0-E(\hbox{\bf p})]},
\label{pole}
\end{equation}
Here
$M^*$ is the in-medium nucleon mass including the scalar part of the self
energy and  $\lambda^*$ is the
nucleon coupling. 
Then one writes the three independent sum rules, one for each Dirac structure
\begin{equation}
-(1-{\xi\over2})\int dp\,{w(p)\over 
2E[p^0-
E]}\simeq{(1-\xi)\over\lambda^{*2}M^*}\int dp\, w(p)\left[
\Pi^{(0,s)}_{OPE}(p)-\Pi^{(0,s)}_{cont}(p)\right]
\end{equation}
\begin{equation}
-(1+{\xi\over2})\lambda^{*2}\int dp\,{w(p)\over 2E
[p^0-E]}\simeq\int dp\, w(p)\left[\Pi^{(0,p)}_{OPE}(p)
-\Pi^{(0,p)}_{cont}(p)\right],\label{sumrules}
\end{equation}
\begin{equation}
-(1+{\xi\over2})\lambda^{*2}V\int dp\,{w(p)\over 2E
[p^0-E]}\simeq\int dp\, w(p)\left[\Pi^{(0,u)}_{OPE}(p)
-\Pi^{(0,u)}_{cont}(p)\right].
\end{equation}
Taking the ratio of these sum rules one can get the needed cancellation
in the effective mass and vector self energy and bring the in-medium
nucleon QCD sum rules in an agreement with the chiral symmetry constraints. 
 Let's consider now the in-medium correlation function of
the vector currents.
The lowest pole of the isovector-vector correlator corresponds 
to the $\rho$-meson
contribution.
Due to relatively large width it decays inside nuclear interior
so the spectrum of the produced dileptons can carry
the information about the modifications of the 
 $\rho$-meson mass and width in matter. Such modification may be related
with partial restoration of CS.One possible  way to look
at the phenomena of CS restoration is to study the 
 correlators describing the in-medium dynamics of the chiral partners.
 The correlators of the chiral partners,
in our case the correlators of the vector and axial-vector currents, should
become identical in the chirally restored phase. Thus one can expect that
these correlators get mixed when the symmetry is only partially restored.
The  effect of chiral mixing indeed takes place both at finite
 temperatures \cite{De}
and densities \cite{Kr}. Making
 use of the standard commutation relation of current algebra
$\left[Q_5^a, J_\nu^b\right]=i\epsilon^{abc}A^{c}_\nu$
and putting it in the  expression for the correlator of the vector currents
$\Pi_V$
one can get 
\begin{equation}
\Pi_V=\Pi^{0}_V+\xi(\Pi^{0}_V-\Pi^{0}_A);
 \Pi_A=\Pi^{0}_A+\xi(\Pi^{0}_A-\Pi^{0}_V)
\label{PiV}
\end{equation}
The parameter $\xi$ in this equation is 4/3 times the one for
the nucleon correlator.
  $\Pi^{0}_V$($\Pi^{0}_A$) is
the correlator of the vector (axial) currents
calculated in the approximation of the noninteracting nucleons.
As one can see from the above equations the correlators get mixed when soft
pion contributions  are taken into account. We note that this statement
is model independent and follows solely from CS. CS  implies
that the correlators of the chiral partners acquire, due to mixing, the 
additional singularities.
These singularities may manifest themselves in the, for example, dilepton 
spectrum, produced in the heavy-ion collisions. It means that the spectrum may
 show the additional enhancement at the energy region close to the mass
of the $A_1$ meson, besides that at the mass of   $\rho$ meson.
 The phenomena of mixing suggests
few possible ways of how chiral symmetry restoration occurs. First,
 the lowest singularities
of the both correlators could indeed coincide at the
 point of CS restoration. Second, the  correlators  may exhibit
two poles of the same strength corresponding to the  $\rho$ and $A_1$ mesons.
Third, the width of the mesons could become large enough to make 
no longer sensible the whole concept of
individual quantum state at high densities.
One notes that the $\omega$ meson being an isospin singlet is not affected
by the chiral corrections. It follows from the fact that the commutator
of the isoscalar vector current with the axial charge $Q_a^5$ is zero.
Let's briefly consider the mixing of the other type of chiral
partners, namely the $\sigma$-$\pi$ pair.  Making use of the
 current algebra commutation relation
$
[Q_5^a, \pi^b]=-\delta^{ab}i\sigma$ and $[Q_5^a, \sigma]=-i\pi^a
$
one can get 
\begin{equation}
\Pi_\pi=(1+\xi)\Pi^{0}_\pi-\xi\Pi^{0}_\sigma;
 \Pi_\sigma=(1+\xi)\Pi^{0}_\sigma-\xi\Pi^{0}_\pi
\label{PiPi}
\end{equation}
Here  the mixing parameter is two times smaller than that for  $\rho$-$A_1$ system and
thus the effect of $\sigma$-$\pi$ mixing is less pronounced at the normal 
nuclear density than in the case of $\rho$-$A_1$ mixing. The point of the complete
restoration of chiral symmetry should, of course, be the same for all kinds of
the chiral partners but the ``velocity'' of approaching to this point may well be 
different. Similar to the case of the $\rho$-$A_1$ system
 the mixing in the 
pseudoscalar-scalar channel practically means that the correlators exhibit the singularities
which are dictated by CS and should be taken into account regardless of the model used to
describe the concrete hadronic processes.  
However, the effect of the $\sigma$-$\pi$ mixing can probably be observed at 
relatively large densities. The case worth studying is deeply bound pion states
\cite {Kk} in heavy nuclei.

\section*{Acknowledgments}

Author would like  to thank Mike Birse for discussions about
the role of chiral symmetry in nuclei.
It is also a pleasure to thank the organizers of Baryons98 for providing 
 with the opportunity
to attend the conference.

\end{document}